\documentclass[fleqn]{article}
\usepackage{espcrc2}
\usepackage{graphicx}
\usepackage{psfig}
\newcommand{\be}{\begin{enumerate}}
\newcommand{\ee}{\end{enumerate}}
\newcommand{\bi}{\begin{itemize}}
\newcommand{\ei}{\end{itemize}}
\newcommand{\jpsi}{J/\psi}

\newcommand{\jpsito}{J/\psi\rightarrow}
\newcommand{\psip}{\psi(2S)}

\newcommand{\pip}{\pi^+}
\newcommand{\pin}{\pi^-}

\newcommand{\ar}{\rightarrow}

\begin{document}
\title{ First observation
of $\jpsi$ and $\psip$ decaying to $n K^0_S\bar\Lambda +c.c.$}
\author{\small{
M.~Ablikim$^{1}$,              J.~Z.~Bai$^{1}$, Y.~Ban$^{13}$,
X.~Cai$^{1}$,                  H.~F.~Chen$^{18}$,
H.~S.~Chen$^{1}$,              H.~X.~Chen$^{1}$, J.~C.~Chen$^{1}$,
Jin~Chen$^{1}$,                Y.~B.~Chen$^{1}$
% ...
\\(BES Collaboration)\\
$^{1}$ Institute of High Energy Physics, Beijing 100049, People's
Republic
of China\\
% ...
$^{13}$ Peking University, Beijing 100871, People's Republic of China\\
% ...
$^{18}$ University of Science and Technology of China, Hefei 230026,
People's Republic of China\\
% ...
}}

\date{\today}

\begin{abstract}
The decays of $\jpsi$ and $\psip$ to
${n}{K^0_S}\bar{\Lambda}+c.c.$ are observed and measured for the
first time, and the perturbative QCD ``12\%'' rule is tested,
based on $5.8 \times 10^7$~$\jpsi$ and $1.4 \times 10^7$~$\psip$
events collected with BESII detector at the Beijing
Electron-Positron Collider. No obvious enhancement near
$n\bar{\Lambda}$ threshold in $\jpsi \to
{n}{K^0_S}\bar{\Lambda}+c.c.$ is observed, and the upper limit on
the branching ratio of $\jpsi \to {K^0_S} X, X \to n \bar \Lambda$
is determined.
\end{abstract}

\maketitle

\section{Introduction}   \label{introd}

Since the discovery of the $\jpsi$ at Brookhaven~\cite{brookhaven} and
SLAC~\cite{slac} in 1974, more than one hundred exclusive decay modes
of the $\jpsi$ have been reported. According to Ref.~\cite{physrep},
direct hadronic, electromagnetic and radiative decays make up
roughly 65\%, 14\%, and 7\% of the total $\jpsi$ decay width,
respectively. However, the measured hadronic decay channels sum up to
less than 35 \%. The BESII data sample of 5.8 $\times 10^7$ $\jpsi$ events
provides a good opportunity to search for missing $J/\psi$ hadronic
decays.

In 2004, BESII reported the observation of an enhancement $X(2075)$
near the threshold of the invariant mass spectrum of
${p}\bar{\Lambda}$ in $\jpsi\ar {p}{K^-}\bar{\Lambda}$ decays. The
mass, width, and product branching fraction of this enhancement are $M
= 2075 \pm 12~({\rm stat.}) \pm 5~({\rm syst.})$ MeV/$c^2$, $\Gamma = 90
\pm 35~({\rm stat.}) \pm 9~({\rm syst.})$ MeV/$c^2$~\cite{pkl}, and
$B(J/\psi \ar K^- X)B(X \ar p \bar{\Lambda}+c.c.) = (5.9 \pm 1.4 \pm 2.0)
\times 10^{-5}$, respectively.  The study of the isospin conjugate
channel $\jpsi\ar n K^0_S \bar{\Lambda}$ is therefore important not
only in exploring new decay modes of $J/\psi$ but also in
understanding the $X(2075)$.

The $5.8 \times 10^7$ $\jpsi$ and $1.4 \times 10^7$ $\psip$ events
at BESII also offer a unique opportunity to search for new decay
modes of $\jpsi$ and $\psip$ and test the ``12\% rule'' in
hadronic decays.  In perturbative QCD, hadronic decays of the
$\jpsi$ and $\psip$ are expected to proceed dominantly via three
gluons or a single direct photon with widths proportional to the
square of the $c\bar c$ wave function at the origin, which is well
determined from dilepton decays.  Thus for any hadronic final
state $h$, the $\jpsi$ and $\psip$ decay branching fractions
should satisfy the so called ``12\% rule''~\cite{rule}.
$$Q_{h}=\frac{B(\psip\ar h)}{B(\jpsi\ar h)}\simeq \frac{B(\psip\ar
e^+ e^-)}{B(\jpsi\ar e^+ e^-)}\simeq 12\%.$$
The leptonic
branching fractions are taken from the particle data group
(PDG)~\cite{pdg2006} tables. It is roughly obeyed for a number of
exclusive hadronic decay channels except some $VP$, $PP$ and $VT$
channels~\cite{rule1}\cite{rule2}\cite{rule3}, where $P, V$ and
$T$ denote members of the pseudoscalar, vector and tensor nonets,
respectively.

In this paper, the first observation and measurement of $\jpsi$ and
$\psip$ to ${n}{K^0_S}\bar{\Lambda}+c.c.$, as well as a test of the
perturbative QCD 12\% rule are presented.  The $\Lambda^*$ and $N^*$
resonance structures in $\jpsi \to {n}{K^0_S}\bar{\Lambda}+c.c.$ are
also shown, where no obvious enhancement near $n\bar{\Lambda}$
threshold is observed.  The upper limit on the branching fraction of
$\jpsi \to {K^0_S} X, X \to n \bar \Lambda$ is determined.

\section{The BES Detector}  \label{BESD}

The upgraded Beijing Spectrometer detector (BESII) is located at the
Beijing Electron-Positron Collider (BEPC). BESII is a large
solid-angle magnetic spectrometer which is described in detail in
Ref.~\cite{besii}.  The momentum of charged particles is determined by
a 40-layer cylindrical main drift chamber (MDC) which has a momentum
resolution of $\sigma_{p}$/p=$1.78\%\sqrt{1+p^2}$ ($p$ in GeV/$c$).
Particle identification is accomplished using specific ionization
($dE/dx$) measurements in the drift chamber and time-of-flight (TOF)
information in a barrel-like array of 48 scintillation counters. The
$dE/dx$ resolution is $\sigma_{dE/dx}\simeq8.0\%$; the TOF resolution
for Bhabha events is $\sigma_{TOF}= 180$ ps.  Radially outside of the
time-of-flight counters is a 12-radiation-length barrel shower counter
(BSC) comprised of gas tubes interleaved with lead sheets. The BSC
measures the energy and direction of photons with resolutions of
$\sigma_{E}/E\simeq21\%\sqrt{E}$ ($E$ in GeV), $\sigma_{\phi}=7.9$
mrad, and $\sigma_{z}=2.3$ cm. The iron flux return of the magnet is
instrumented with three double layers of proportional counters that
are used to identify muons.

A GEANT3 based Monte Carlo (MC) program (SIMBES)~\cite{pid} with
detailed consideration of the detector performance is used.  The
consistency between data and MC has been carefully checked in many
high purity physics channels, and the agreement is reasonable.  More
details on this comparison can be found in Ref.~\cite{pid}. The
detection efficiency and mass resolution for each decay mode in this
analysis are obtained from MC simulation.

\section{Analysis}\label{analysis}

The analyzed $J/\psi$ and $\psi(2S)\ar {n}{K^0_S}\bar{\Lambda}$ with
${K^0_S} \to \pi^+\pi^-$ and $\bar{\Lambda} \to \bar p \pi^+$ (and
$c.c.$) final states contain four charged tracks and an undetected
neutron or anti-neutron.  We require the candidate events to satisfy
the following common selection criteria:

\begin{enumerate}
\item Events must have four good charged tracks with zero net charge.
  A good charged track is a track that is well fitted to a
  three-dimensional helix, originates from the interaction region and
  has a polar angle $\theta$ in the range $|\cos \theta| <$ 0.8.
  Because of the long decay lengths before ${K^0_S}$ and
  $\bar{\Lambda}$ decay to $\pi^+\pi^-$ and $\bar p \pi^+$, the
  interaction region is defined as R$_{xy} <$ 0.12 m and $|z| <$ 0.3 m.
  Here, R$_{xy}$ is the distance from the beamline to the point of
  closest approach of the track to the beamline, and $|z| $ is the
  distance along the beamline to this point from the interaction
  point.

\item For each charged track in an event, $\chi^{2}_{PID}(i)$ is
  determined using both $dE/dx$ and TOF information:
\begin{center}
 $\chi^{2}_{PID}(i)$ = $\chi^{2}_{dE/dx}(i) + \chi^{2}_{TOF}(i),$
\end{center}
where $i$ corresponds to the particle hypothesis.  A charged track is
identified as a $p$ if $\chi^{2}_{PID}$ for the $p$ hypothesis is less
than those for the $\pi$ or $K$ hypotheses.  For the channels studied,
one charged track must be identified as a $p$ or $\bar{p}$.
\end{enumerate}

\subsection{\boldmath Measurement of $\jpsi\ar n K^0_S \bar\Lambda +c.c. $}
\label{jpsinkl}

For $\jpsi\ar {n}{K^0_S}\bar{\Lambda}\ar \bar{p} n \pip\pin\pip$, the
$K_S^0 \ar \pi^+ \pi^-$ and $\bar{\Lambda} \ar \bar{p}\pi^+$ decays are
reconstructed using secondary vertex fitting, and the $\pi^+$ from the
$\bar{\Lambda}$ decay is identified.  To select $\Lambda
(\bar{\Lambda})$ [$\Lambda (\bar{\Lambda})$ mass selection],
$|M_{p\pi^-(\bar p\pi^+)}-1.115|<0.012$ GeV/c$^2$ is required, and to
select $K^0_S$ ($K^0_S$ mass selection), $|M_{\pi^+ \pi^-}-0.497|<0.02$
GeV/c$^2$ is required.  To reject the backgrounds from channels
containing a $K^0_S$ but no $\Lambda$, like e.g. $\jpsi\ar
\bar{p}{K^0_S}\Sigma^{-}+c.c.$, we require $L_{xy}(\Lambda)$, the
distance from the reconstructed $\Lambda$ vertex to the event origin,
to be larger than 5 mm.

\begin{figure}[htbp]
\centerline{\hbox{\psfig{file=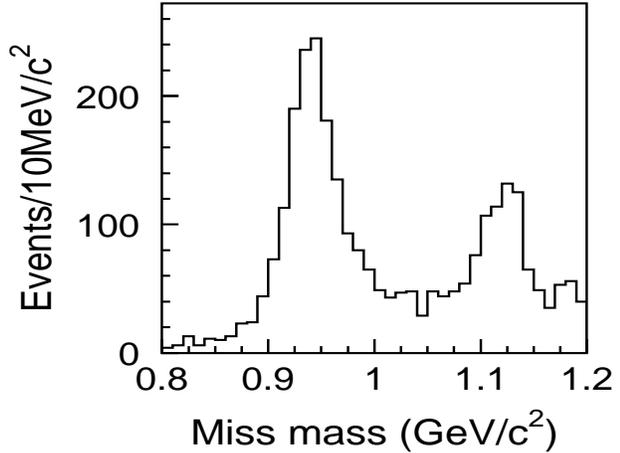,width=8cm,height=6cm}}}
\caption{The missing mass determined from the charged tracks in $\jpsi\ar n
  K^0_S\bar\Lambda+c.c.$ candidate events satisfying $\Lambda
  (\bar{\Lambda})$ and $K^0_S$ mass selections and $L_{xy}(\Lambda) > 5$ mm.}
\label{recoil}
\end{figure}

\begin{figure}[htbp]
\centerline{\hbox{\psfig{file=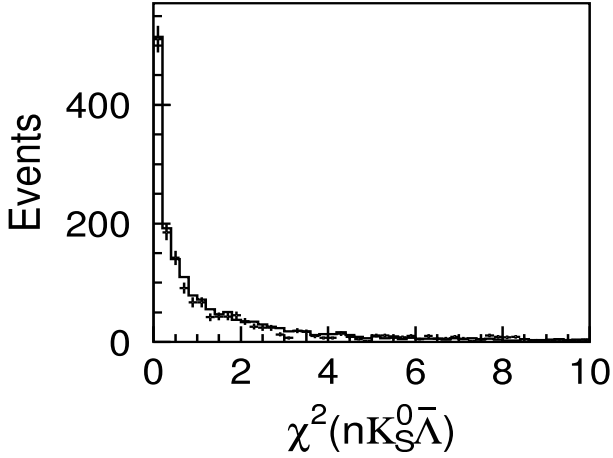,width=8cm,height=6cm}}}
\caption{The $\chi^2$ distributions for the 1C fits to
  the $J/\psi \to \bar p n \pip\pin\pip$ hypothesis for $J/\psi \to n
  K^0_S \bar\Lambda$ candidate events for the selection used in
  Fig~\ref{recoil}. The crosses are data; the full histograms are the
  sum of MC simulation of $\jpsi\ar n K^0_S \bar\Lambda$ and
  background determined from $K^0_S$ sidebands ($0.06 < |M(\pi^+
  \pi^-) - 0.497| < 0.08$ GeV/c$^2$).}
\label{chisq}
\end{figure}

Figure~\ref{recoil} is the missing mass spectrum determined from the
charged tracks in $J/\psi \to n K_S^0 \bar{\Lambda} + c.c.$ candidate
events satisfying $\Lambda (\bar{\Lambda})$ and $K^0_S$ mass
selections and $L_{xy}(\Lambda) > 5$ mm. A clear peak at the nominal
neutron mass is observed. The second peak in the high missing mass
region comes from $J/\psi\ar n K^0_S
\bar{\Sigma^0}(1385)+c.c.$ and
$J/\psi\ar\Sigma^-\bar\Sigma^+(1385)+c.c.$ backgrounds. To suppress
background and improve the resolution, a one constraint (1C) kinematic
fit with a missing neutron is applied under the $J/\psi \to
\bar p n \pi^+\pi^-\pi^+$ hypothesis.  The distribution of 1C fit
$\chi^{2}_{\bar p n \pip\pin\pip}$ for the above selection is shown in
Fig.~\ref{chisq}. The agreement between data and MC simulation is
reasonable, and in the following, $\chi_{1C}^2 < 5$ is required.

\begin{figure}[htbp]
  \centerline{\hbox{\psfig{file=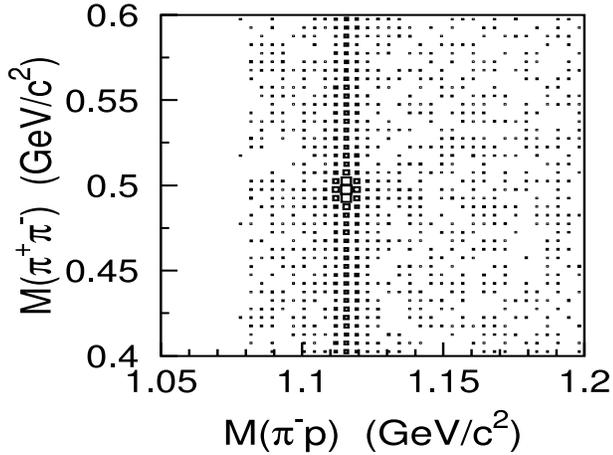,width=8cm,height=6cm}}}
\caption{Scatter plot of $m_{\bar{p}\pip}$ versus $m_{\pip\pin}$ for
  $\jpsi\ar n K^0_S\bar\Lambda $ candidate events satisfying
  $\chi_{1C}^2 < 5$ and $L_{xy}(\Lambda) > 5$ mm.}
\label{plot}
\end{figure}

\begin{figure}[htpb]
\centerline{\hbox{\psfig{file=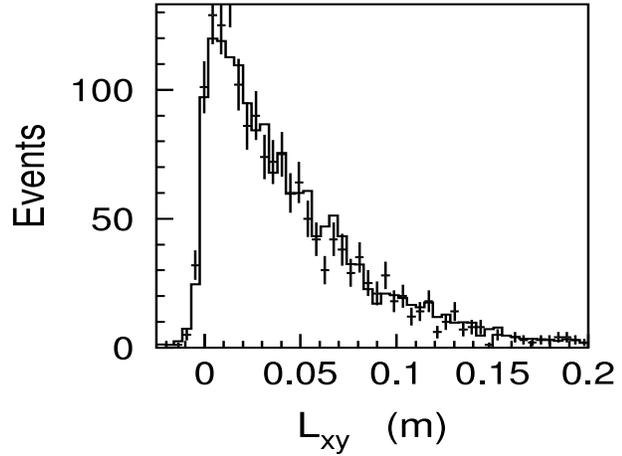,width=8cm,height=6cm}}}
\caption{ The  $\Lambda(\bar{\Lambda})$ decay length distributions
  with the $\Lambda (\bar{\Lambda})$ and $K^0_S$ mass selection
  requirements and $\chi_{1C}^2 < 5$ for data and MC. The histogram is
  the sum of signal MC and background from $K^0_S$ sidebands, and the
  crosses are data.}
\label{lxy}
\end{figure}

\begin{figure}[htbp]
\centerline{\hbox{\psfig{file=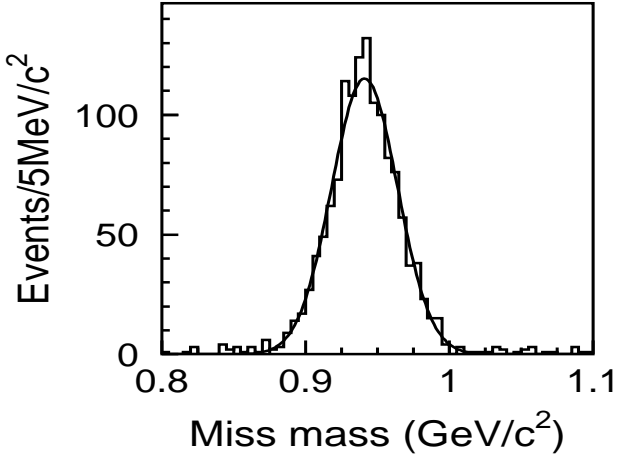,width=8cm,height=6cm}}}
\caption{The missing mass spectrum of charged tracks in $\jpsi\ar n
K^0_S\bar\Lambda+c.c.$ for events satisfying the requirements in
Fig.~\ref{lxy} and $L_{xy}(\Lambda) > 5$ mm.}
\label{missingmass}
\end{figure}

\begin{figure}[htbp]
\centerline{\hbox{\psfig{file=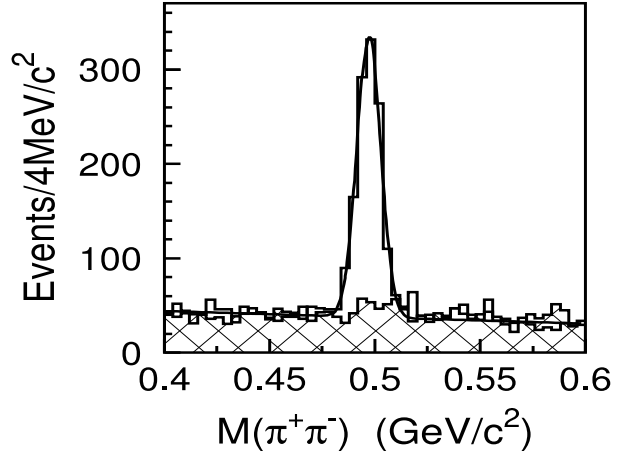,width=8cm,height=6cm}}}
\caption{The $\pip\pin$ invariant mass distribution for $\jpsi\ar n
  K^0_S\bar\Lambda+c.c.$ candidates satisfying the $\Lambda
  (\bar{\Lambda})$ mass selection requirements, $\chi_{1C}^2 < 5$, and
  $L_{xy}(\Lambda) > 5$ mm. The fit is also shown. The cross-hatched
  area is the sum of the backgrounds after normalization, as described
  in the text.}
\label{fitsig}
\end{figure}

Figure~\ref{plot} shows the scatter plot of $m_{\bar{p}\pip}$ versus
$m_{\pip\pin}$, and clear $\bar{\Lambda}$ and $K^0_S$ signals are
seen.  Figure~\ref{lxy} shows the $\Lambda$ decay length distributions
for data and MC simulation for events satisfying $\Lambda(\bar{\Lambda})$
and $K^0_S$ mass selection requirements and $\chi_{1C}^2 < 5$.
The missing mass distribution of charged tracks for events satisfying
these requirments plus $L_{xy}(\Lambda) > 5$ mm is shown in 
Fig.~\ref{missingmass}, and a very clean neutron peak is seen. A fit
with a Gaussian function yields a mass value consistent with that of
the neutron. The $\pi^+\pi^-$ invariant mass spectrum is shown in
Fig.~\ref{fitsig}, and a $K_S^0$ signal is clearly seen.

Backgrounds for $\jpsi\ar n K^0_S\bar\Lambda$ which contribute to the
peak in the $K^0_S$ signal region mainly come from $\jpsi\ar n
K^0_S\bar\Sigma^0 $ and $J/\psi\rightarrow{p}{K^0_S}\bar\Sigma^-$ that
survive selection criteria. Normalizing with the corresponding
branching fractions and the number of $J/\psi$ events in the data sample,
a total of $42\pm 8$ $\jpsi\ar n K^0_S\bar\Sigma^0 $ and $12 \pm 3$
$J/\psi\rightarrow{p}{K^0_S}\bar\Sigma^-$ background events are
esitmated.  These events will be subtracted in determining the final
branching fractions. Other surviving background events mainly come from
$\jpsito$ $\bar{\Lambda}{\Sigma^-}\pi^+$,
$\bar{\Lambda}{\Sigma^+}\pi^-$, ${\Sigma^+}{\bar\Sigma^-(1385)}$, and
${\Sigma^-}\bar\Sigma^+(1385)$, and their charge conjugate channels,
but they only give a flat contribution in the $K^0_S$ signal region.
The sum of these backgrounds, normalized by their branching fractions,
is shown as the cross-hatched area in Fig.~\ref{fitsig}, and it is
consistent with the background under the peak for data.

Using a Gaussian to describe the $K^0_S$ and a second order polynomial
function to model the background shape, a fit to the $m_{\pip\pin}$
distribution is performed, shown as the curve in Fig.~\ref{fitsig}. A
total of 1058$\pm$33 $K^0_S$ events are obtained. No $K^0_S$ signal is
observed in the $m_{\pip\pin}$ invariant mass distribution for events
which recoil against the $\bar\Lambda$ sideband region
($1.140<m_{p\pi}<1.164$ GeV/$c^2$).  The detection efficiency for the
signal is 6.09\%, which is determined from a uniform phase space
MC simulation.  The branching fraction is:
$$B(\jpsi\ar{n}{K^0_S}\bar{\Lambda}+c.c.)$$
$$=\frac{N_{obs}-N_{bg}}{N_{\jpsi} \cdot \epsilon\cdot B(\bar{\Lambda}\ar
\bar{p}\pip) \cdot B(K^0_S\ar \pip\pin)}$$
$$=(6.46\pm0.20)\times 10^{-4},$$
where $N_{obs}$ is the number of events
observed (1058$\pm$33); $N_{bg}$ is the number of background events from
$\jpsi\ar n K^0_S\bar\Sigma^0 $ (42 $\pm$ 8) and ${p}{K^0_S}\bar\Sigma^-$
(12 $\pm$ 3); $\epsilon$ is the detection efficiency; $N_{\jpsi}$ is
the number of $\jpsi$ events; and $B(\bar{\Lambda}\ar \bar{p}\pip)$ and
$B(K^0_S\ar \pip\pin)$ are the branching fractions
of $\bar{\Lambda}\ar \bar{p}\pip $ and $K^0_S\ar \pip\pin$~\cite{pdg2006}.
The error is statistical only.

If we fit the charge conjugate channels separately, we obtain
$502\pm22$ events with an efficiency of $6.02\%$ for
$J/\psi\rightarrow {n}{K^0_S}\bar{\Lambda}$, $560\pm24$ events, an
efficiency of $6.16\%$ for $J/\psi\rightarrow
{\bar{n}}{K^0_S}{\Lambda}$, and the following branching fractions:
\begin{center}
 $B(\jpsi\ar{n}{K^0_S}\bar{\Lambda})=(3.09\pm 0.14)\times 10^{-4}$,\\
 $B(\jpsi\ar\bar{n}{K^0_S}\Lambda)=(3.39\pm 0.15)\times 10^{-4}$,
\end{center}
where the errors are statistical only. These results are consistent
with each other in $1.5\sigma$.

\begin{figure}[htbp]
\centerline{\hbox{\psfig{file=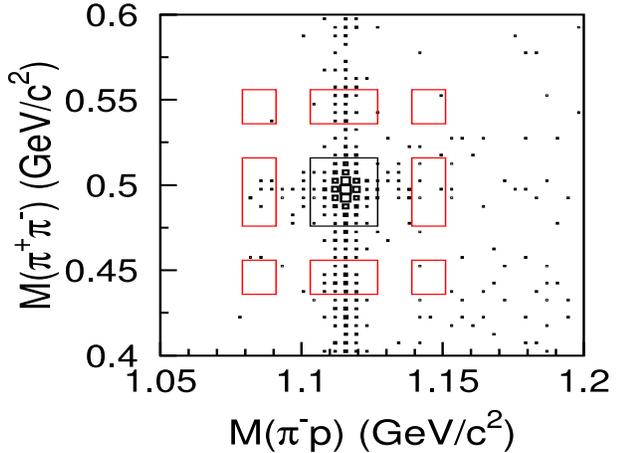,width=8cm,height=6cm}}}
\caption{The scatter plot of $m_{\pin {p}}$ versus $m_{\pip\pin}$ for
  $\jpsi\ar n K^0_S\bar\Lambda$ candidates after all selection
  requirements except for the $\Lambda$ and $K^0_S$ mass requirements.
  The boxes in the plot show the signal and sideband regions.}
\label{scatter-f}
\end{figure}

\begin{figure}[htbp]
\centerline{\hbox{\psfig{file=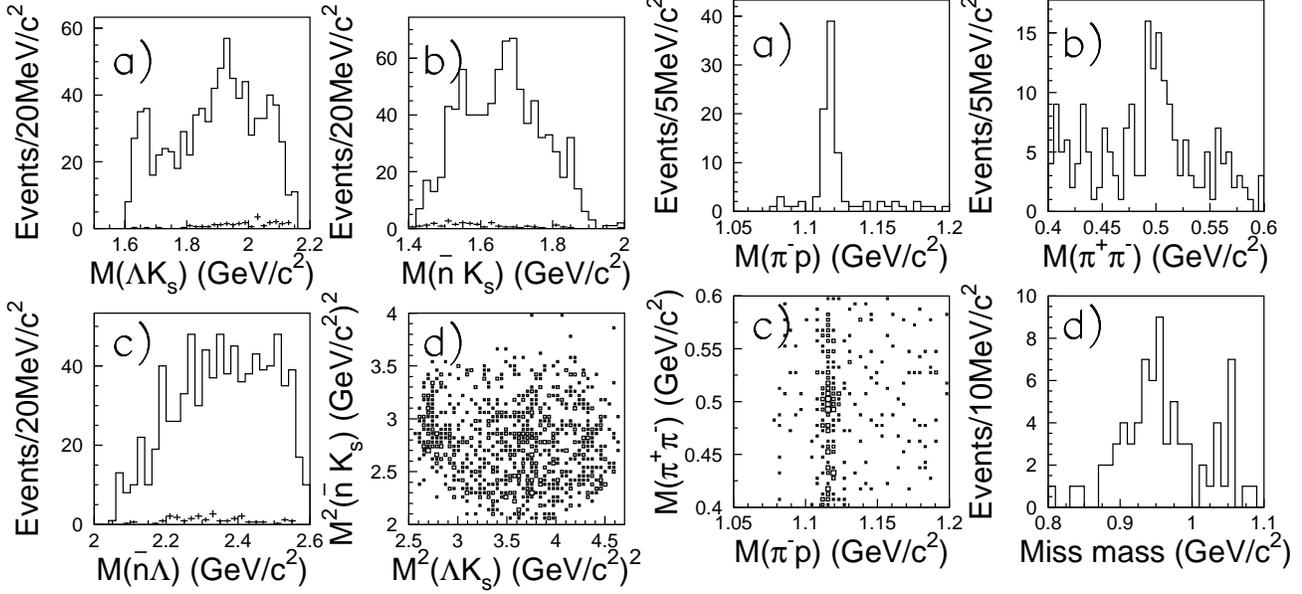,width=9cm,height=9cm}}}
\caption{The invariant mass spectra of (a) ${\Lambda K_S^0}$, (b) ${\bar n K_S^0}$,
  and (c) $\bar {n} \Lambda$, as well as (d)
  the Dalitz plot for candidate events after all selection criteria.
  The crosses show the sideband backgrounds.}
\label{Nstar}
\end{figure}

In order to obtain a clean sample of $\jpsi\ar n K^0_S
\bar{\Lambda}$ and $\bar{n}K^0_S \Lambda$, we require events to
satisfy the $\Lambda (\bar{\Lambda})$ and $K^0_S$ mass selection
requirements, $\chi_{1C}^2 < 5$, and $L_{xy}(\Lambda) > 5$ mm, and
also require the $K^0_S$ decay length $L_{xy}(K^0_S) >5$ mm to
eliminate backgrounds without a $K^0_S$ in the final state, such
as $\jpsi\ar\bar{\Lambda}\Sigma^-\pip$.  After this final
selection, the background contribution is estimated to be less
than 5\%.  Figure~\ref{scatter-f} shows the scatter plot of
$m_{\pin p}$ versus $m_{\pip\pin}$ for $\jpsi\ar n
K^0_S\bar\Lambda $ candidate events for all but the $\Lambda
(\bar{\Lambda})$ and $K^0_S$ mass selection requirements, where
the boxes in the plot show the signal and sideband regions. The
invariant mass spectra of ${\Lambda K_S^0}$, ${n K_S^0}$, and
${\bar {\Lambda} n (\Lambda \bar n)}$, as well as the Dalitz plot
for all selection requirements are shown in Fig.~\ref{Nstar}.  In
the $\Lambda K_S^0$ invariant mass spectrum, an enhancement near
$\Lambda K_S^0$ threshold is evident, as is found in the $\Lambda
K$ mass spectrum in $J/\psi \to p K^- \bar{\Lambda}$~\cite{pkl2}.
If the enhancement is fitted with an acceptance weighted S-wave
Breit-Wigner function and a function $f_{bg}(\delta)$ describing
the phase space ``background'' contribution, the fit leads to
M=1.648$\pm$0.006GeV/$c^2$ and $\Gamma=61\pm21$MeV/$c^2$,
respectively. Here the errors are only statistical. The systematic
uncertainties are not included since more accurate measurements of
the mass and width should come from a full PWA involving
interferences among $N^*$ and $\Lambda^*$ states. The fitted mass
and width are consistent with those obtained from a partial wave
analysis of $J/\psi \to p K^- \bar{\Lambda}$~\cite{pkl2}. The
$X(2075)$ signal which was seen in the $p\bar{\Lambda}$ invariant
mass spectrum in $\jpsito p K^- \bar{\Lambda}$ is not significant
here.  Using a Bayesian approach~\cite{Bayesian} and fixing the
mass and width of $X(2075)$ to 2075 MeV/$c^2$ and 90 MeV/$c^2$
respectively, the upper limit on the number of events observed
$N_{obs}^{UL}$ is 54 events at the 90\% C.L.

The $N^*$ state at around 1.9 GeV/c$^2$ in the $\Lambda K_S^0$
invariant mass spectrum and the $\Lambda^*$ states at around 1.5
and 1.7 GeV/c$^2$ in the $n K^0_S$ invariant mass spectrum are
present. A larger data sample and a partial wave analysis are
needed to analyze these $N^*$ and $\Lambda^*$ states.

\vspace{0.5cm}
\subsection{\boldmath Measurement of $\psip\ar n K^0_S \bar\Lambda + c.c.$ }
\label{psipnkl}
\begin{figure}[htbp]
\centerline{\hbox{\psfig{file=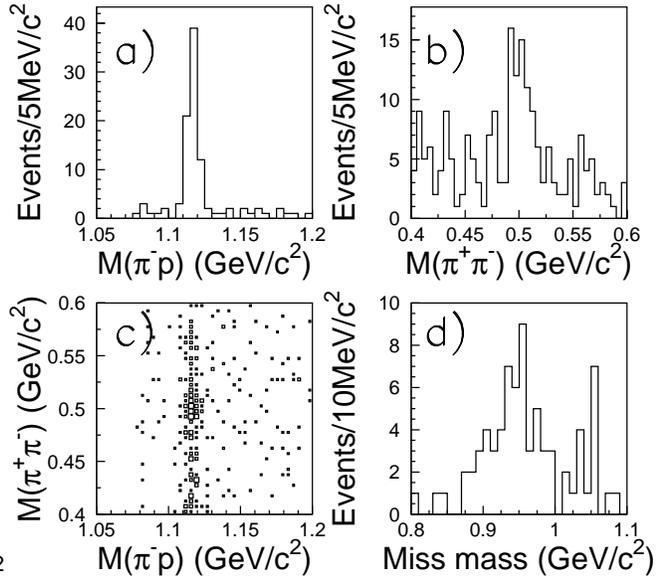,width=9cm,height=9cm}}}
\caption{The invariant mass spectra of (a) ${\pin p}$ and (b)
  ${\pi^+\pi^-}$, (c)
  the scatter plot of $M_{\pin p}$ versus $M_{\pi^+\pi^-}$, and (d) the
  missing mass spectrum for events satisfying the $\Lambda
  (\bar{\Lambda})$ mass selection requirement, $\chi_{1C}^2 < 5$, and
  $L_{xy}(\Lambda) > 5$ mm.}
\label{psipplot}
\end{figure}

Using the same criteria as in Section
\ref{jpsinkl}, we select $\psip\ar {n}{K^0_S}\bar{\Lambda}+c.c.$
events from the BESII sample of 14M $\psip$ events. The $\pin p$
and $\pi^+\pi^-$ invariant mass spectra, the scatter plot of $m_{
\pin p}$ versus $m_{\pi^+\pi^-}$, and the missing mass spectrum
after the final selection are shown in Fig.~\ref{psipplot}. The
$\Lambda$ and $K_S^0$ signals are obvious.

\begin{figure}[htbp]
\centerline{\hbox{\psfig{file=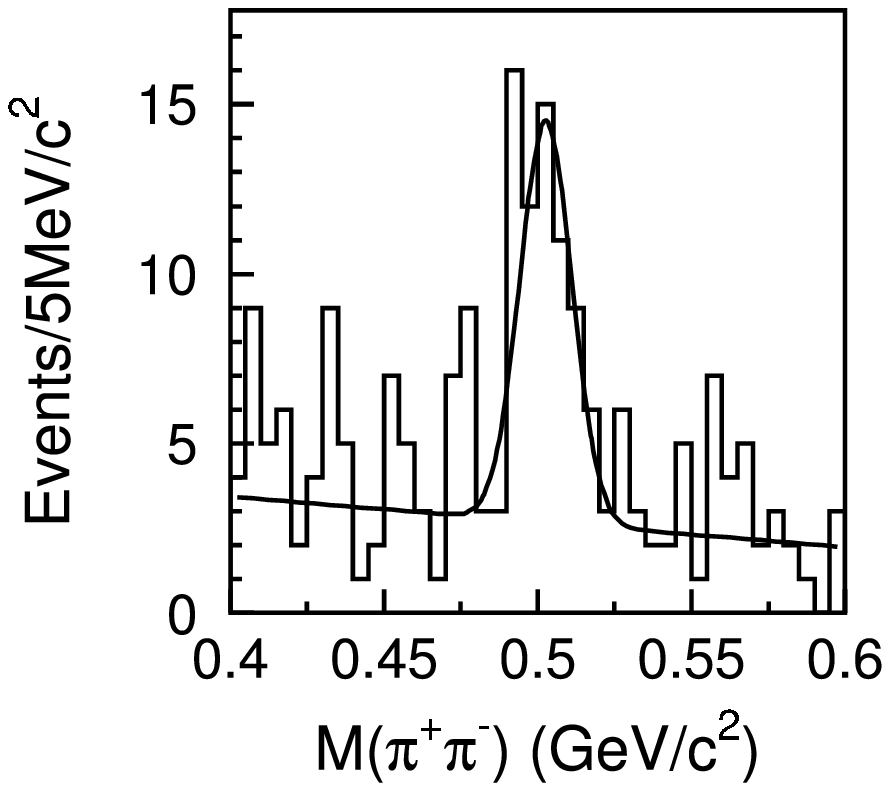,width=8.0cm,height=6.0cm}}}
\caption{The fit of $\pi^+\pi^-$ invariant mass spectrum for
$\psip\ar n K^0_S \bar\Lambda +c.c.$ for events satisfying the
requirements in Fig.~\ref{psipplot}d.}
\label{psipks}
\end{figure}

Fitting the $\pip\pin$ mass spectrum with a Gaussian for the $K^0_S$ signal
and a first order polynomial background, as shown in
Fig.~\ref{psipks}, yields 50$\pm$7 $K_S^0$ events. The statistical
significance of the
$K^0_S$ is about 7.2$\sigma$. The $2\pm 1$ background events
from the $\Lambda(\bar{\Lambda})$ sidebands and 2$\pm$1
background events from $\psip\ar n K^0_S \bar\Sigma^0$ are subtracted.
A uniform phase space MC simulation determines the detection
efficiency to be 9.16\%. The
corresponding branching fraction is:
\begin{center}
$B(\psip\ar {n}{K^0_S}\bar{\Lambda}+c.c.)=(8.11\pm 1.14)\times 10^{-5}$.
\end{center}
Here the error is statistical only.

\vspace{0.5cm}
\subsection{Systematic Errors} \label{J-sys}
In this analysis, the systematic errors on the branching fractions
mainly come from following sources.

\subsubsection{MDC tracking}
The MDC tracking efficiency has been measured using channels like
$J/\psi \to \rho \pi$, $\Lambda \bar{\Lambda}$, and
$\psi(2S)\ar\pip\pin\jpsi,\jpsi\ar\mu^+\mu^-$. The MC simulation agrees with data within 1 to 2\% for each charged
track \cite{pid}. Thus 8\% is taken as the systematic error coming
from MDC tracking for the 4-prong events considered in this analysis.

\subsubsection{Kinematic fit}
The systematic error from the 1C kinematic fit should be smaller than that
from the 4C kinematic fit, since there are fewer constraints. Various studies
show that the uncertainty of the 4C kinematic fit is
around 4\%~\cite{shencp}.
Here we conservatively take 4\% as the error from the 1C kinematic
fit.

\subsubsection{Particle identification}

In Ref.~\cite{pid}, the particle identification
efficiency of $\pi$, $K$, and $p$ are analyzed in detail. Here, only one
charged track is required to be identified as a $p$ or $\bar{p}$, and
the systematic error from particle identification is less than 2\%.

\subsubsection{$\Lambda$ and $K^0_S$ vertex finding}

In Ref.~\cite{jiaojb},
$\jpsi\ar\Lambda\bar{\Lambda}\ar\pip\pin{p}\bar{p}$ is chosen as the
reference channel to study the systematic error of the $\Lambda$
vertex finding algorithm, and 1.2\% is determined as the systematic error
for one $\Lambda$ vertex. For $K^0_S$, the efficiency of the secondary
vertex finding is studied using $J/\psi\ar K^*(892)\bar K+c.c.$
events, and the systematic error is about 4.1\% ~\cite{wangzhe}.

\subsubsection{MC model}

Different hadronization models for simulating the hadronic
interactions give different detection efficiencies. Their
differences are taken as systematic errors. The systematic errors
are  7.0\%  and 14.7\% for
$J/\psi\rightarrow{n}{K^0_S}\bar{\Lambda}$ and its conjugate
channel, respectively, and  11.1\% for $\psip\rightarrow{n}
{K^0_S}\bar{\Lambda}+c.c.$. The efficiency differences with or
without considering the intermediate $N^*$ and $\Lambda^*$ states
are also taken as the systematic errors. They are 5.3\% and 4.5\%
for $J/\psi\rightarrow{n}{K^0_S}\bar{\Lambda}$ and
$J/\psi\rightarrow\bar{n}{K^0_S}{\Lambda}$, respectively.

\subsubsection{Background uncertainty}

The background uncertainties come from the uncertainties
associated with the estimation of the sideband backgrounds, continuum
events, and the events from other background channels, as well as the
uncertainties of the background shape, different fit ranges,
etc. Therefore, the statistical errors in the estimated background
events, the largest difference in changing the background shape, and
the difference of changing the fit ranges are taken as the systematic
errors for the background uncertainty.

\subsubsection{Intermediate decay branching fractions}

The branching fractions of $\Lambda\ar{p}\pin$ and the
$K^0_S\ar\pip\pin$ decays are taken from the PDG~\cite{pdg2006}. The
errors on these branching fractions are taken as systematic errors in
our measurements.

\subsubsection{Number of $\jpsi$ and $\psip$ events}

The total number of $\jpsi$ events is (57.70$\pm$ 2.62)$\times
10^6$, determined from inclusive 4-prong hadrons~\cite{fangss},
and the total number of $\psi(2S)$ events $N_{\psip}$ is
(14.0$\pm$0.6)$\times 10^6$, determined from inclusive hadronic
events~\cite{moxh}.  The uncertainty on the number of $\jpsi$
events, 4.7\%, and the uncertainty on the number of $\psip$
events, 4.0\%, are also systematic errors.

The above systematic errors are all listed in Table~\ref{syserr}.
The total systematic error is determined by adding all
terms in quadrature.

\section{Results}
 The decays of $\jpsi$ and $\psip$ to $n K^0_S\bar\Lambda + c.c.$  are
observed for the first time, and their branching fractions are:
\begin{center}
$B(\jpsi\ar n K^0_S\bar\Lambda+c.c.)=(6.46\pm0.20\pm1.07)\times 10^{-4}$,\\
$B(\jpsi\ar n K^0_S\bar\Lambda)=(3.09\pm0.14\pm0.58)\times 10^{-4}$,\\
$B(\jpsi\ar \bar n K^0_S\Lambda)=(3.39\pm0.15\pm0.48)\times 10^{-4}$,\\
$B(\psip\ar n K^0_S\bar\Lambda+c.c.)=(0.81\pm0.11\pm0.14)\times 10^{-4}$.
\end{center}
The ratio of the branching fractions
of $\psip$ and $\jpsi$ decaying to $n K^0_S\bar{\Lambda}+c.c.$ is
$Q_{h}$ = (12.6 $\pm$ 3.5)\% and obeys the 12\% rule well.

There is no obvious enhancement near $n\bar\Lambda$ threshold. The
upper limit on the branching fraction on the near-threshold
enhancement $X(2075)$ at $n\bar\Lambda$ threshold at the 90 \%
C.L. is:
$$B(\jpsi \ar {K^0_S} X) \cdot B( X \ar {n} \bar{\Lambda}+c.c.)$$
$$=\frac{N_{obs}^{UL}}{N_{\jpsi}\cdot \epsilon \cdot B(\Lambda\ar p\pin)\cdot
B(K^0_S\ar \pip\pin)\cdot (1 - \delta_{sys})}$$
$$<4.9\times 10^{-5}~ (90 \% ~C.L.),$$
where $N_{obs}^{UL}$ is 54 events; $\epsilon$=5.32\% is the detection
efficiency considering the angular distributions;
$N_{\jpsi}$ is the number of $\jpsi$ events;
$B(\Lambda\ar p\pin)$ and $B(K^0_S\ar \pip\pin)$ are the  $\Lambda\ar
p\pin$ and $K^0_S\ar \pip\pin$ branching fractions, and $\delta_{sys}$
is the systematic error (17.3\%).
Taking into account the isospin factor, the branching fraction upper
limit for
$B(\jpsi \ar {K^0_S} X) \cdot B( X \ar {n} \bar{\Lambda}+c.c.)$
is not inconsistent with that for
$B(\jpsi \ar K X) \cdot B( X \ar {p} \bar{\Lambda}+c.c.)$~\cite{pkl}.

\begin{table*}[!h]
\caption{Summary of the systematic errors.}
\begin{center}
\begin{tabular}{c|c|c|c|c}
\hline
\hline
Sources &    \multicolumn{3}{|c}{Relative Error (\%)}\\
\hline
decay modes & $\jpsi\ar{n}{K^0_S}\bar{\Lambda}$ &
$\jpsi\ar\bar{n}{K^0_S}{\Lambda}$ &
$\jpsi\ar X(2075){K^0_S}$ & $\psip\ar{n}{K^0_S}\bar{\Lambda}+c.c.$  \\
\hline

MDC tracking & 8.0 & 8.0 & 8.0 & 8.0 \\

Kinematic fit & 4.0 & 4.0 & 4.0 & 4.0 \\

Particle ID   & 2.0 & 2.0 & 2.0 & 2.0 \\

MC model &  15.6 & 8.3 & 9.4 & 11.1 \\

Background uncertainty & 2.1 & 5.1 & 9.3 & 7.1 \\

$\Lambda$ and $K^0_S$ reconstruction & 1.2 & 1.2 & 4.3 & 1.2 \\

Intermediate decay branching fractions& 0.9  & 0.9 & 0.9 & 0.9\\

Numbers of $\jpsi$ and $\psip$ events  & 4.7 & 4.7 & 4.7 & 4.0 \\
\hline
Total systematic error & 18.9 & 14.3 & 17.3 &  16.6 \\
\hline
\hline
\end{tabular}
\end{center}
\label{syserr}
\end{table*}

\section{Acknowledgments}

The BES collaboration thanks the staff of BEPC and computing
center for their hard
efforts. This work is supported in part by the National Natural
Science Foundation of China under contracts Nos. 10491300,
10225524, 10225525, 10425523, 10625524, 10521003, the Chinese Academy
of Sciences under contract No. KJ 95T-03, the 100 Talents Program of
CAS under Contract Nos. U-11, U-24, U-25, and the Knowledge Innovation
Project of CAS under Contract Nos. U-602, U-34 (IHEP), the
National Natural Science Foundation of China under Contract No.
10225522 (Tsinghua University), and the Department of Energy under
Contract No. DE-FG02-04ER41291 (U. Hawaii).



\begin{thebibliography}{xx}
\bibitem{brookhaven} J. J. Aubert, {\em et al.}, Phys. Rev. Lett. {\bf 33},1404 (1974).
\bibitem{slac} J. E. Augustin, {\em et al.}, Phys. Rev. Lett. {\bf 33},1406 (1974).
\bibitem{physrep} L. K\"opke and A. Wermes, Phys. Rep. {\bf 174}, 67 (1989).
\bibitem{pkl} BES Collaboration, M. Ablikim {\em et al.}, Phys. Rev. Lett.
{\bf 93},
112002 (2004).
\bibitem{rule} T. Appelquist and H. D. Politzer, Phys, Rev. Lett.{\bf 34}, 43
(1975); A. De Rujula and S. L. Glashow, {\it ibid}, page 46.
\bibitem{pdg2006} W.-M. Yao {\em et al.}~(Particle Data
Group), J. Phys. G {\bf 33},1 (2006).
\bibitem{rule1} P. B Mackenzie, G. P. Lepage, Phys. Rev. Lett. {\bf 47}, 1244 (1981).
\bibitem{rule2} W. S. Hou and A. Soni, Phys. Rev. lett. {\bf 50}, 569 (1983).
\bibitem{rule3} W. S. Hou, Phys. Rev. D {\bf 55}, 6952 (1992).
\bibitem{besii} BES Collaboration, J. Z. Bai {\em et al.}, Nucl. Instrum.
and Methods A {\bf 458}, 627 (2001).
\bibitem{pid}  BES Collaboration, M. Ablikim {\em et al.}, Nucl. Instrum. and
Methods A {\bf 552}, 344 (2005).
\bibitem{pkl2} H. X. Yang for BES Collaboration,
Int. J. Mod. Phys. A {\bf 20}, 1985 (2005).
\bibitem{Bayesian} G. J Feldman and R. D. Cousins, Phys. Rev. D {\bf 57},3873
(1998).
\bibitem{shencp} BES Collaboration, M. Ablikim {\em et al.}, Phys. Rev. D
{\bf 74}, 012004 (2006).
\bibitem{jiaojb} BES Collaboration, M. Ablikim {\em et al.}, Phys. Lett.
B {\bf 648}, 149 (2007).
\bibitem{wangzhe} BES Collaboration, M. Ablikim {\em et al.}, Phys. Rev.
Lett. {\bf 92}, 052001 (2004).
\bibitem{fangss} S. S. Fang {\it et al.}, HEP\&NP {\bf 27}, 277 (2003) (in Chinese).
\bibitem{moxh} X. H. Mo {\it et al.}, HEP\&NP, {\bf 28}, 455 (2004).
\end{thebibliography}
\end{document}